\documentclass[aps,prl,reprint,groupedaddress,showpacs]{revtex4-1}
\usepackage{latexsym,amssymb,amsfonts,mathrsfs,color,graphicx,amsmath}
\usepackage[commentmarkup=uwave]{changes}
\begin{document}
	
\title{Chirality-induced bacterial rheotaxis in bulk shear flows}

\author{Guangyin Jing,$^{1,2,\dagger}$}
\author{Andreas Z\"{o}ttl$^{2,3,\dagger}$}
\email{andreas.zoettl@tuwien.ac.at}
\author{\'Eric Cl\'ement$^{2}$}
\author{Anke Lindner$^{2}$}
\date{\today}

\affiliation{$^{1}$School of Physics, Northwest University, Xi'An, 710127, China}
\affiliation{$^{2}$Physique et M\'ecanique des Milieux H\'et\`erog\`enes, PMMH, ESPCI Paris, PSL University,CNRS, Sorbonne Universit\'e, Université de Paris, 10, rue Vauquelin, 75005 Paris, France}
\affiliation{$^{3}$Institute for Theoretical Physics, TU Wien, Wiedner Hauptstra{\ss}e 8-10, Wien, Austria}
\affiliation{$^\dagger$ These authors contributed equally to this work}

\begin{abstract}
  Interaction of swimming bacteria with flows controls their ability to explore complex environments,  crucial to many societal and environmental challenges and relevant for microfluidic applications  as cell sorting. Combining experimental, numerical and theoretical analysis, we present a comprehensive study of the transport of motile bacteria in shear flows. Experimentally, we obtain with high accuracy and for a large range of flow rates, the spatially resolved velocity and orientation distributions. They are in excellent agreement with the simulations of a kinematic model accounting for stochastic and microhydrodynamic properties and in particular the flagella chirality. Theoretical analysis reveals the scaling laws behind the average rheotactic velocity at moderate shear rates using a chirality parameter and explains the reorientation dynamics leading to a saturation at large shear rates from the marginal stability of a fixed point.  Our findings constitute a full understanding of the physical mechanisms and relevant parameters of bacteria bulk rheotaxis.
\end{abstract}

% insert suggested PACS numbers in braces on next line
% insert suggested keywords - APS authors don't need to do this
%\keywords{}
%\maketitle must follow title, authors, abstract, \pacs, and \keywords
\maketitle

\section*{Introduction}

\begin{figure*}[tb]
  \centering
  \includegraphics[width=.7\textwidth]{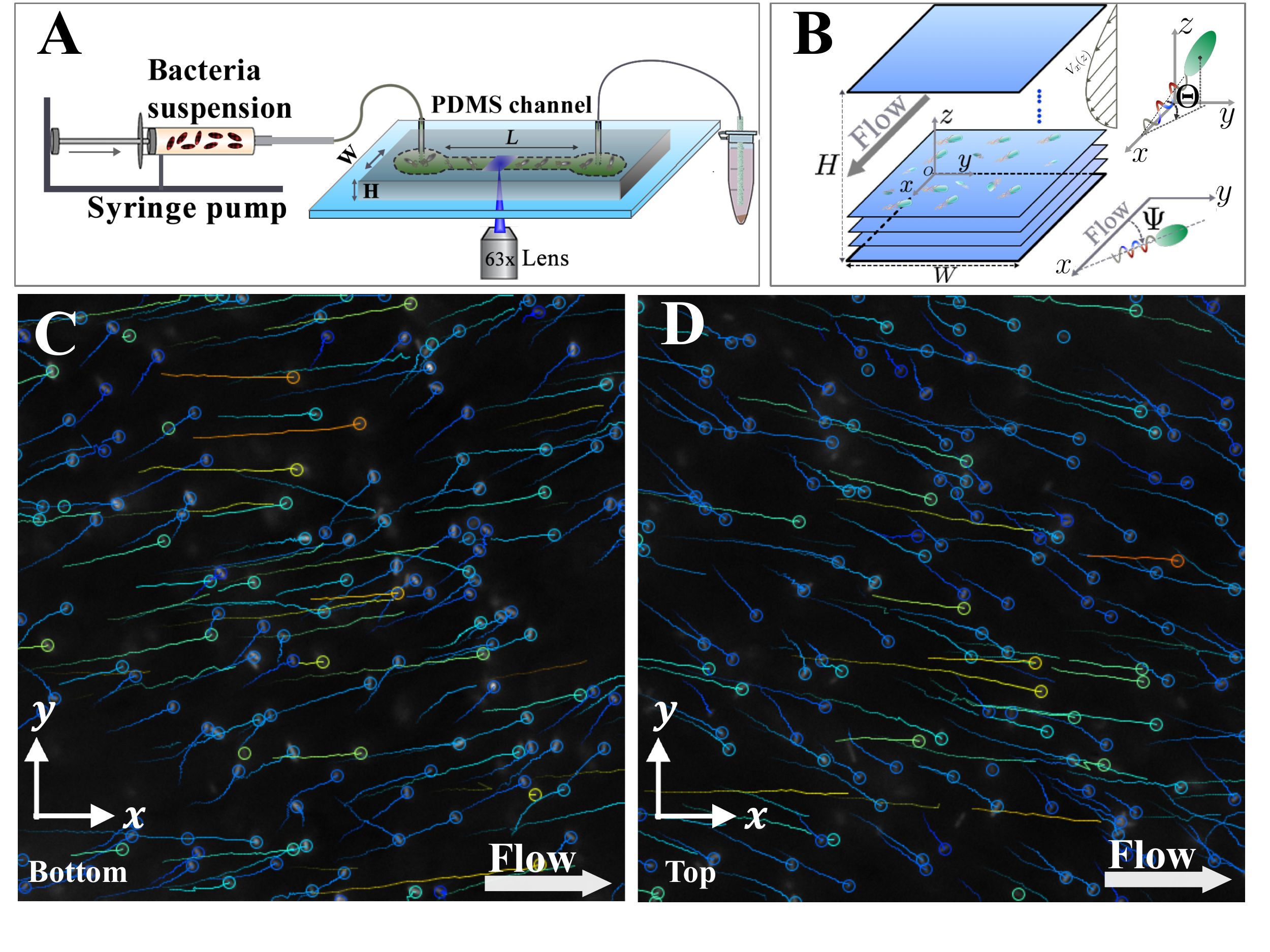}\\
  \caption{\label{SetupTopBottomTracks} \textbf{Setup and typical trajectories of swimming bacteria in the upper and lower half of the channel.}
    \textbf{(\textit{A})} A dilute bacterial suspension is injected into a PDMS microchannel (width $W=600 \mu m$, height $H=100 \mu m$, and up to $L=20 mm$ in length)  at a given flow rate $Q$. \textbf{(\textit{B})} Bacteria and passive tracers are recorded at 200 fps using a 63X lens (observation window in the $x,y$ plane $200 \mu m \times 100 \mu m$) at varying distances $z$ from the bottom wall. The angle $\Psi$ defines the bacteria orientation in the $x,y$ plane and $\Theta$ is the out-of-plane angle. \textbf{(\textit{C,D})} Typical trajectories of swimming bacteria in the lower (C) and upper channel half. (D). Bacteria drift towards the right with respect to the negative flow direction in  \textbf{(\textit{C})} and in the opposite direction in  \textbf{(\textit{D})}.}
\end{figure*}

The interaction of swimming microorganisms with flows determines their ability to move in complex environments such as biological channels, soils or medical conducts. The understanding of the resulting dynamics is crucial for a number of societal and environmental challenges, as infections, soil purification, contamination of biomedical devices, but also cell sorting and analysis \cite{costerton1999bacterial,von2005infections,bain2014global,hatzenpichler2016visualizing}. Particularly interesting are situations where microorganisms do not just follow the local flow velocity to be transported downstream along stream lines, but orient with respect to the flow and show preferential transport up- or downstream or even a side-wise drift. This non-trivial organization under flow is expected to be at the origin of transport anomalies observed in the dispersion process in capillary tubes or porous media, which remains today poorly understood \cite{Chilukuri2016, Alonso2019, Creppy2019}. From a fundamental point of view these transport dynamics are  determined by the microorganism shape, activity and rotational (or run-and-tumble) noise in combination with the given flow properties. 

Passive non-chiral rigid particles transported in viscous flows generally follow streamlines, even if complex orientation dynamics can be present as a function of the particle shape \cite{Bretherton1962}. For example, elongated passive objects in shear flows perform so-called Jeffery orbits \cite{jeffery1922motion, Bretherton1962}, periodically changing their orientation while transported downstream along stream lines. Such orbits have been observed in different experimental systems with and without Brownian noise \cite{Liu2018, Einarsson2016, Zoettl2019} and the role of fluctuations on the orbits has been addressed theoretically \cite{Zoettl2019}.

For passive particles drifts along velocity gradients have only been observed in more complex situations and in the presence of shear gradients, for example in viscoelastic flows \cite{DAvino2017}, in the presence of inertia \cite{DiCarlo2009} or for flexible particles \cite{duRoure2019, Kumar2012}. 
When particle symmetry is broken by chirality, particles can migrate towards the vorticity directions \emph{i.e.}\ perpendicular to velocity gradients, as has been predicted and experimentally observed \cite{makino2005migration, MarcosPRL2009}. Whether a drift towards the right/left (positive/negative vorticity direction)
is observed depends on the handedness of the particle and the sign of the local shear rate. For all these systems drift velocities remain small compared to flow velocities and their influence only becomes noticeable after long distances.

This changes fundamentally when particles become active. For motile microorganisms orientation dynamics, mainly governed by Jefferey dynamics \cite{jeffery1922motion}, directly translate into swimming directions and drift velocities become of the order of swimming velocities.
In shear flow microswimmers crossing streamlines lead to new families of ``active-Jeffery-orbits''.  In Poiseuille flow ``swinging'' and ``shear-tumbling'' trajectories \cite{ZottlPRL2012, Zottl2013} were identified theoretically and numerically, and their existence was recently confirmed experimentally for motile \textit{E.~coli} bacteria \cite{Junot2019}. 
  Moreover in Poiseuille flow, kinetic  theory \cite{Ezhilan2015} predicts that the interplay between stochastic reorientation, active swimming and the varying local shear rate leads to preferred upstream and downstream swimming.

Microorganism transport has primarily been studied close to solid surfaces, mainly due to the fact that activity leads to surface accumulation \cite{Zottl2016}
In addition, in these regions transport velocities are small while large velocity gradients, in combination with specific surface interactions, strongly influence microswimmer dynamics. It has been found half a century ago  \cite{Bretherton1961} that microorganisms can orient with respect to flow gradients to surfaces.  Upstream orientation has been observed for sperm cells \cite{Bretherton1961, Kantsler2014, Tung2015}, for \textit{E.~coli} bacteria \cite{Hill2007, Kaya2012, FigueroaSoftmatter2015, Altshuler2013} and artificial microswimmers \cite{Palacci2015, ren2017rheotaxis, baker2019fight, brosseau2019relating}. This upstream motion has been analysed theoretically  \cite{Costanzo2012, Uspal2015, baker2019fight, brosseau2019relating} and is generally attributed to fore-aft asymmetry of the swimmer shape. In addition, organisms reorient on average towards the positive vorticity direction \cite{Hill2007, Kaya2012, Marcos2012, FigueroaSoftmatter2015}, an effect attributed to the counterrotation of cell body and flagella.
At high enough shear rates the interplay of different effects can lead to oscillatory surface rheotaxis \cite{Mathijssen2019}.

In the bulk, bacterial rheotaxis has been discovered by Marcos {\it et al.}~\cite{Marcos2012} who have shown that surface interactions are not required to observe bacteria drift towards the vorticity direction  but that a helical flagella shape leading to chirality-induced lift forces was sufficient. In combination with the viscous drag on the bacteria head this leads to a rheotactic torque \cite{Marcos2012} reorienting bacteria in flow resulting in
  a net rheotactic drift which has 
  the opposite sign compared to the drift of passive helices \cite{MarcosPRL2009},  and has been experimentally observed for swimming \textit{bacillus subtilis} at a distance $H/4$ away from the walls of a microchannel Poiseuille flow \cite{Marcos2012}.
  The rheotactic drift has been characterized in the form of a mean rheotactic velocity as a function of the local shear rate and was quantitatively reproduced using a hydrodynamic model  of a microswimmer based on resistive force theory in combination with Brownian noise \cite{Marcos2012}.
Despite the fact that the numerical model presented by Marcos {\it et al.} \cite{Marcos2012}
  reproduces quantitatively, by parametric adjustment, the mean rheotactic drift
  velocity, the details of the underlying physical mechanisms were not fully revealed.

Here we combine an experimental, numerical and theoretical analysis to systematically investigate bulk rheotaxis of a wild type {\it E.~coli} bacteria. Microfluidic experiments are performed in wide channels and a large number of bacteria tracks are recorded as projections into planes parallel to the bottom wall at different distances from the latter (Fig.~\ref{SetupTopBottomTracks}A,B). Careful tracking of active bacteria as well as passive tracers allows then to obtain precise rheotactic velocity and orientation distributions at different positions of the Poiseuille flow at various flow rates.
To match quantitatively the experimental results, we propose an extended description of the swimming kinematics reflecting the chiral nature of the propelling flagella in addition to the standard Jeffery dynamics and rotational and run and tumble noise. 
The combined role of the drag on the bacterial body and the flagellar chiral structure are encoded into a single number, the chiral strength,  reflecting the reorientation effect in response to the local velocity gradients.
An empirical value of this coupling paramenter is determined for a wild type \textit{E.~coli}.

Theoretical analysis reveals the scaling laws behind the average rheotactic velocities for a full range or shear rates. We define a dimensionless shear rate in the form of a chirality number combining the effects of chiral strength and rotational P\'eclet number leading to a collapse of all experimental and numerical data sets obtained from a parameter study.
  At low and moderate shear rates,
  we reveal that the rheotactic drift velocity increases linearly with the chirality number. At larger chirality number, from a linear stability analysis around a marginally stable fixed point, we explain the reorientation dynamics leading to a very slow approach of the rheocatic drift towards the swimming velocity value.
Our findings constitute a full characterization of bacteria rheotaxis under shear flow, spanning the complete space of the Poiseuille channel flow and a comprehensive understanding of the physical mechanisms and relevant parameters behind it.

\section{Results and discussion}
\label{sec:results}

\subsection{Mean rheotactic velocity}
\label{subsec:meanrheotactivvelocity}

\begin{figure*}
  \centering
  \includegraphics[width=0.8\textwidth]{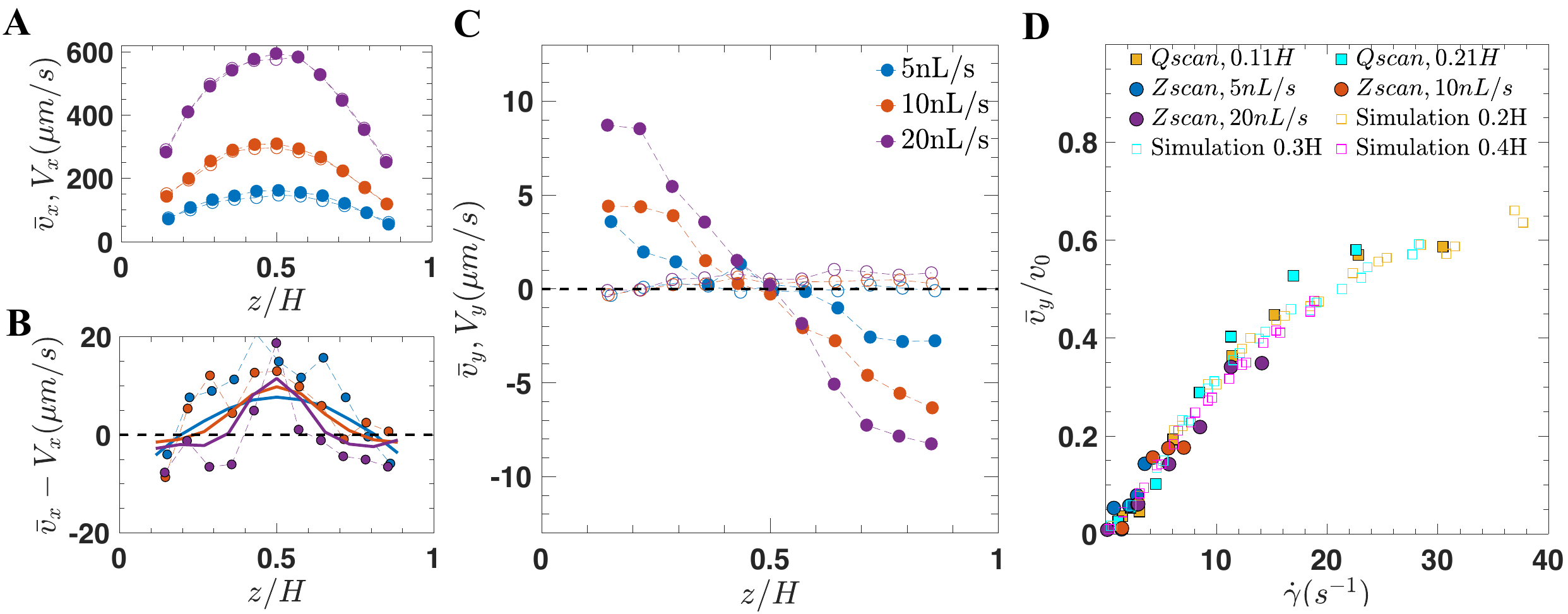}
  \caption{ \label{Fig2:MeanVxVy}
    \textbf{Mean velocities for bacteria and passive tracer beads.}   \textbf{(\textit{A})}. Bacteria ($\bar{v}_x(z)$) and bead ($V_x(z)$) velocities obtained by scanning through the $z$-direction.  Passive tracer beads with diameter $d=1\mu m$ (empty circles) and bacteria (filled circles) are represented at mean flow rates of $Q=5$ (blue), 10 (red), 20 $nL/s$ (purple). \textbf{(\textit{B}) Difference between bacteria and flow velocities $\bar{v}_x-V_x$.} Results from simulations are shown by solid lines. \textbf{(\textit{C})}. Corresponding mean rheotactic velocities $\bar{v}_y$ and bead velocities $V_y(z)$. \textbf{(\textit{D})} Rheotactic velocity $\bar{v}_y$ normalized by the average bacteria swimming speed $v_0$ as a function of local shear rate $\dot{\gamma}$ controlled by two methods: z-scan: scanning through the channel height at given flow rates $Q$ ($5, 10, 20 nL/s$), corresponding to the data of (B); Q-scan: varying flow rates at fixed channel height (0.11H, 0.21H). Results from simulations are indicated by open symbols and are give for heights 0.2H, 0.3H and 0.4H.}
\end{figure*}
	
A dilute suspension of {\it E.~coli} bacteria is injected at a given flow rate $Q$ into a microchannel of width $W$=600$\mu m$ and height $H$=100$\mu m$ (Fig.~\ref{SetupTopBottomTracks}A)  which imposes to a very good approximation a planar Poiseuille flow  $V_x(z) = 4V_{max}z(H-z)/H^2$ sufficiently away from the side lateral walls, with $V_{max}$ the flow velocity in the center of the channel. The local shear rate is then $\dot{\gamma}=\partial V_x(z) /\partial z=4V_{max}(H-2z)/H^2$. Using a high magnification objective lens bacteria tracks are recorded as projections into the $x$-$y$ plane at different distances from the bottom wall (Fig.~\ref{SetupTopBottomTracks}B), and far away ($\gtrsim 250 \mu m$) from lateral side walls. Exact calibration of the local flow velocities and channel orientation is obtained by the simultaneous recording of passive tracers for all experiments performed. This calibration step is essential for the determination of bacteria orientation and velocity distributions.  Since we focus on bulk dynamics bacterial trajectories close to the top and bottom channel walls are not included in our analysis.

Fig.~\ref{SetupTopBottomTracks}C,D show examples of bacterial trajectories in layers in the upper and lower parts of the microchannel Poiseuille flow. One clearly sees a bias towards the  right with respect to the negative flow direction on panel C and towards the left on panel D, confirming the rheotactic cross-streamline migration in the $y$ direction induced by the chirality of the left handed bacteria flagella \cite{Marcos2012} and being a function of the sign of the local shear rate.

Quantitative measurements of mean bacteria velocities ($\bar{v}_x$,$\bar{v}_y$) as well as transport velocities of passive tracers ($V_x$,$V_y$) for different flow rates $Q$ are displayed in Fig.~\ref{Fig2:MeanVxVy}A,C in the flow direction ($x$-direction) and in the vorticity direction ($y$-direction), respectively, as a function of the distance $z$ from the bottom wall. The passive tracers (open symbols) follow a Poiseuille profile ($V_x(z)$ in Fig.~\ref{Fig2:MeanVxVy}A) without a drift in $y$-direction (Fig.~\ref{Fig2:MeanVxVy}C) indicating the perfect alignment of channel and microscope and indeed the $x$-axis is identical to the flow direction.

Different local shear rates $\dot{\gamma}$ can be obtained in two ways, either by varying $V_{max}$ via the imposed flow rate $Q$ or by varying the position $z$ inside the channel. For a systematic analysis of bacterial rheotaxis, allowing to independently investigate the role of local shear rate and positions within the Poiseuille flow, both  $Q$-scans and $z$ scans are performed.

Mean bacteria velocities (filled symbols) deviate from the background velocities both in the parallel ($x$-) direction and in the transverse ($y$-) direction. In the $x$-direction bacteria velocities are higher in the center of the channel and lower close to channel boundaries compared to the background flow.
This is illustrated more clearly in Fig.~\ref{Fig2:MeanVxVy}B where the $z$-dependent relative velocities between bacteria and background flow are shown.
This indicates a preferred orientation of bacteria downstream in the channel center and upstream closer to the channel walls for all considered flow rates.
Mean bacteria  velocities $\bar{v}_y$ confirm the visual observations from Fig.~\ref{SetupTopBottomTracks}C,D and have opposite signs in the lower and upper half of the Poiseuille flow. This rheotactic velocity $\bar{v}_y$ increases with increasing flow rate $Q$ and with decreasing distance $z$ from the wall. Note that wall effects are visible for the data points at a distance of 0.1$H$ ($\sim 10 \mu$m) from top and bottom wall and will be excluded from further analysis in this paper. 

Fig.~\ref{Fig2:MeanVxVy}C shows mean drift velocities $\bar{v}_y$ as a function of local shear rates from different data sets, including the results from the three $z$-scans from Fig.~\ref{Fig2:MeanVxVy}B and from two different $Q$-scans. We note that we scale the velocities by the average bacteria velocity $v_0=25\mu ms^{-1}$ in the absence of flow (see Supplemental Information). Representing the mean velocities as a function of the local shear rate leads to a reasonable data collapse indicating that the local shear rate is the main control parameter of our system. Within our range of shear rates, the increase in mean drift velocity is first linear with local shear rate, similar as  observed in \cite{Marcos2012}, and then reaches a maximum of around half the average bacteria swimming velocity at the highest shear rate.

\subsection{Theoretical framework}
\label{subsec:theory}
	
In order to understand the physical mechanisms behind bacterial rheotaxis in microchannel flow we develop a theoretical framework which captures the dynamics of individual non-interacting bacteria. We account for
the elongated and chiral shape of the flagellated bacteria,
their self-propulsion velocity $v$, tumbling, translational and rotational noise,
and their advection and rotation in Poiseuille flow. Denoting the instantaneous position of a bacterium by $\mathbf{r}$ and its orientation by $\mathbf{e}$, we can write the dynamics of a bacterium in Poiseuille flow as
      
\begin{equation}
  \frac{\text{d}\mathbf{r}}{\text{d}t} = v\mathbf{e} + \mathbf{V}_x(z)  + \mathcal{H}\cdot \boldsymbol{\xi}
  \label{Eq:EOM11a}
\end{equation}
for the positional dynamics and
\begin{equation}
  \frac{\text{d}\mathbf{e}}{\text{d}t} = [\boldsymbol{\Omega}^J + \boldsymbol{\Omega}^C  + \sqrt{2D_r}\boldsymbol{\xi}^r] \times \mathbf{e}
  \label{Eq:EOM11}
\end{equation}
for the orientational dynamics.
For Poiseuille flow these two equations are coupled  and Eq.~(\ref{Eq:EOM11a}) includes
orientation-dependent self-propulsion, advection in Poiseuille flow and anisotropic translational diffusion, 
and Eq.~(\ref{Eq:EOM11}) position-dependent reorientation in flow.
The random numbers $\xi_i$ and $\xi_i^r$ model Gaussian white noise  (see Methods),
$\mathcal{H}$ is related to the anisotropic translational diffusion tensor (see Methods), and $D_r$ is the rotational diffusion constant. We use
the well-known Jeffery reorientation rate $\boldsymbol{\Omega}^J$ to capture the rotation in flow due to elongation \cite{jeffery1922motion, pedley1992hydrodynamic, Zottl2013, Junot2019},

\begin{equation}
  \begin{split}
    \Omega^J_x &= \dot{\gamma} \frac 1 2 Ge_xe_y \\
    \Omega^J_y &= \dot{\gamma} \frac 1 2 \left( 1 + G(e_z^2-e_x^2) \right) \\
    \Omega^J_z &= -\dot{\gamma} \frac 1 2 G e_y e_z
    \label{Eq:OmegaJ}
  \end{split}
\end{equation}
with the Bretherton shape factor $G=(\alpha^2-1)/(\alpha^2+1) \lesssim 1$,
where $\alpha$ is the effective aspect ratio of the bacteria.
The chiral strength $\nu$ of a bacterium is related to the specific shape of the left-handed chiral flagella bundle and the size of the head \cite{Marcos2012,Mathijssen2019} leading to a
chirality-induced reorientation rate \cite{Mathijssen2019} $\boldsymbol{\Omega}^C$
\begin{equation}
  \begin{split}
    \Omega^C_x &= -\nu\dot{\gamma} e_z(2e_x^2-1) \\
    \Omega^C_y &= -2\nu \dot{\gamma} e_xe_ye_z\\
    \Omega^C_z &= -\nu \dot{\gamma} e_x(2e_x^2-1).
    \label{Eq:OmegaC}
  \end{split}
\end{equation}
In order to model tumbling we include tumble events which modify the bacterium orientation instantaneously at exponentially distributed times (see Methods).
We note that we neglect a  term in  Eq.~(\ref{Eq:EOM11a}) which captures a small passive rheotactic drift force but which is negligible compared to the other terms for our regime of parameters.

The theoretical model (Eqs.~(\ref{Eq:EOM11a})-(\ref{Eq:OmegaC})) contains a number of parameters that need to be determined. The swimming speed $v$ of individual bacteria is assumed to be normal-distributed  with mean $v_0=25 \mu m s^{-1}$ and standard deviation of $8 \mu m s^{-1}$ in order to closely match  experimental conditions in the absence of flow (see Supplementary Materials).  While the value of $D_r=0.67s^{-1}$ \cite{drescher2011fluid} and the tumbling statistics \cite{berg1972chemotaxis} can be estimated from the literature, we treat $\alpha$ and $\nu$ as free parameters.
These parameters can be adjusted using the precise experimental results.
We thus perform Brownian dynamics simulations of the coupled equations (\ref{Eq:EOM11a}) and (\ref{Eq:EOM11})  including tumbling and simple steric repulsion for swimmer-wall interactions to be compared to the experimental results for the mean rheotactic velocities $\bar{v}_y$ as a function of the local shear rate $\dot{\gamma}$ as shown in Fig.~\ref{Fig2:MeanVxVy}C.

The initial slope of  $\bar{v}_y(\dot{\gamma})$  can be used to adjust the mean chiral strength of the bacteria to $\nu=0.06$, being of the same order as estimated previously using resistive force theory (RFT) \cite{Marcos2012,Mathijssen2019}. The parameter $\nu$ only depends on the bacterial shape. In particular, its value depends in a non-trivial form on the helical shape of the flagella bundle, is maximized for large cell bodies and vanishes in the limit of very small cell body size. The effective aspect ratio $\alpha$ is determined from the deviation from the linear regime towards the saturation of the rheotactic velocity at high shear rates, with the results not being very sensitive to the value of the parameter $\alpha$. Here we use $\alpha=5$, in agreement with typical experimental values \cite{Junot2019}. These values for $\alpha$ and $\nu$ are used for all further comparison between experiments and simulations.

From Fig.~\ref{Fig2:MeanVxVy}D we can see that adjusting these parameters the numerical model indeed describes very well the experimental observations. As in the experiments, data obtained from different positions $z$ in the channel and different flow strength collapse onto almost a single curve, but small deviations stem from the velocities obtained from different layers in the channel.

In  Fig.~\ref{Fig2:MeanVxVy}B we compare the longitudinal velocity differences between bacteria and the Poiseuille flow from numerical simulations with the experiments,  which indeed show the same results, \textit{i.e.}\ a preferred upstream orientation near the walls and a downstream orientation in the center. This is in excellent agreement with a kinematic model which neglects the swimmer chirality \cite{Ezhilan2015}. This indicates that chirality does not influence, at least qualitatively,   the polarization profile along the flow direction but mainly affects the bacteria orientations in the transverse direction.

\begin{figure*}
  \centering
  \includegraphics[width=0.7\textwidth]{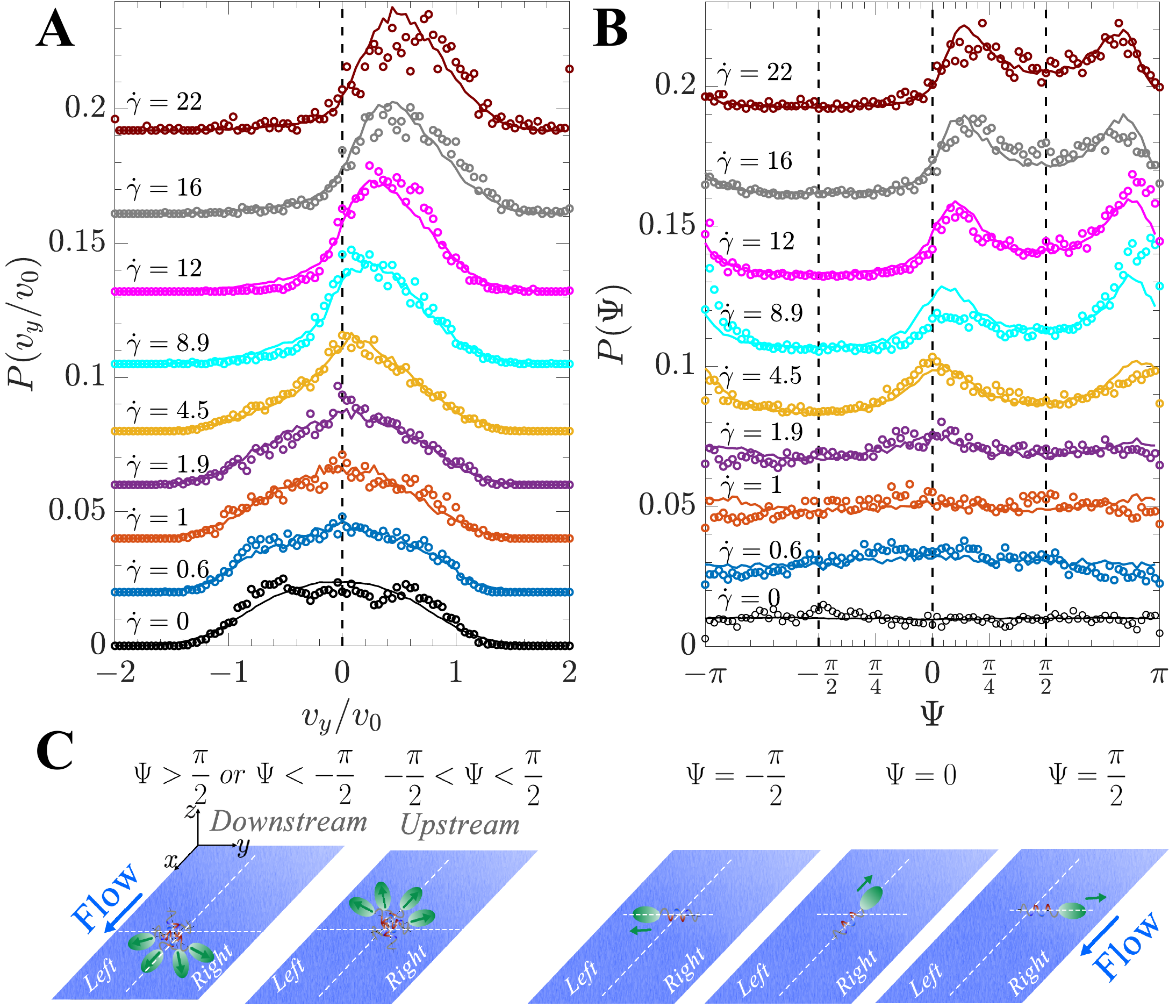}
  \caption{ \textbf{Velocity and orientation distributions.}
    Experimental (symbols) and numerical (solid lines) results of \textbf{(\textit{A})} velocity $v_y$ and \textbf{(\textit{B})}  orientation $\Psi$ distributions obtained by varying the flow rate $Q$ at a given distance from the channel bottom wall ($z\approx H/4$). Local shear rates have been closely matched between experiments and simulations. For better readability the different curves are shifted in the vertical direction. \textbf{(\textit{C}) Sketch of the bacterium orientation $\Psi$ and relation to up- and downstream, and left and right orientation.} }
  \label{Fig3:ExpNumVyPsiDistribution}
\end{figure*}

\begin{figure*}[bth]
  \centering
  \includegraphics[width=0.8\textwidth]{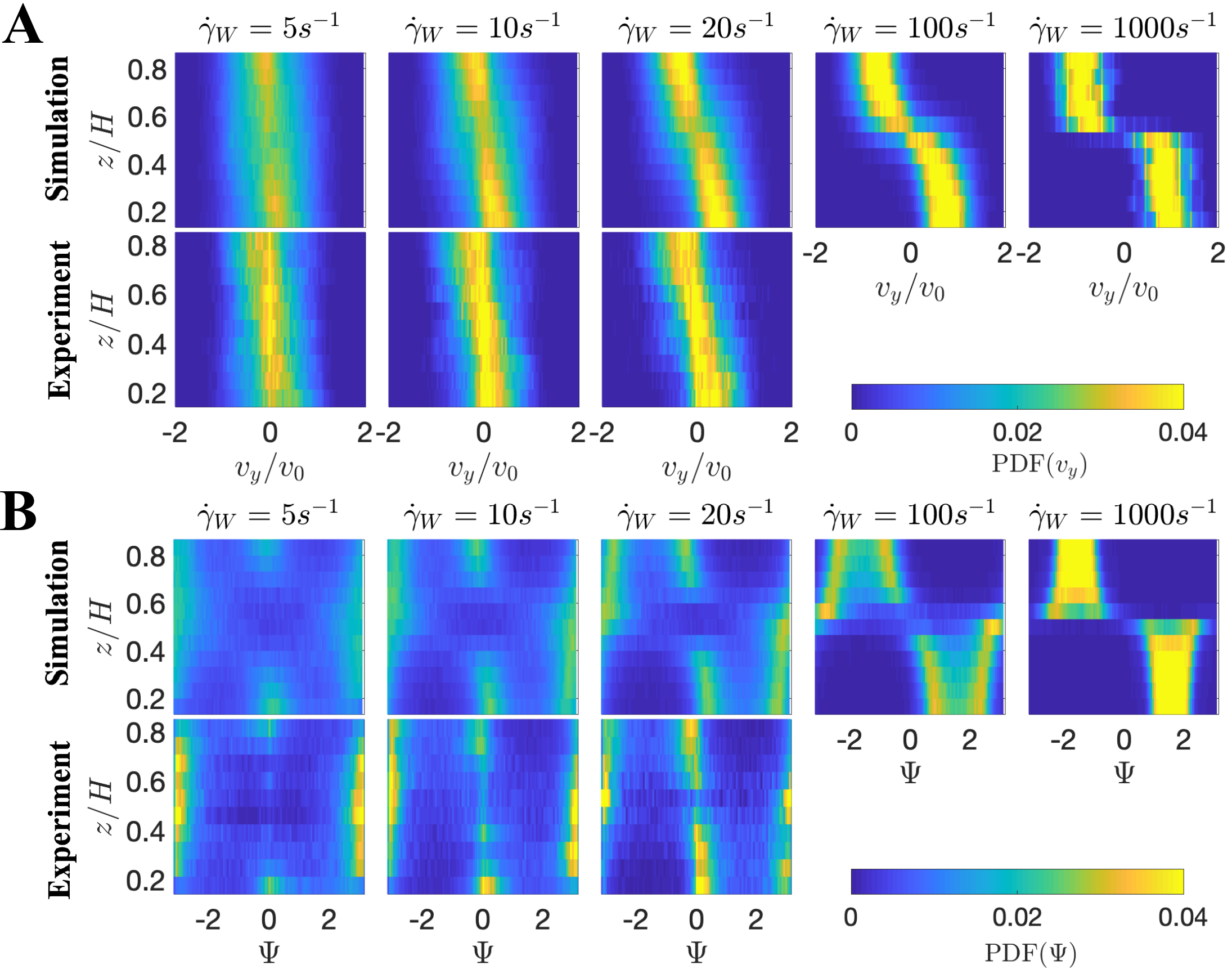}
  \caption{ \textbf{Color map of rheotactic velocity and orientation distributions as a function of channel height  from experiments and simulations.} Different panels correspond to different applied flow rates, as indicated by the corresponding wall shear rates $\dot{\gamma}_w$. 	
  }
  \label{Fig4:distributionschannelheight}
\end{figure*}

\subsection{Velocity and orientation distributions}
\label{subsec:distributions}

We now turn to velocity and orientation distributions. Fig.~\ref{Fig3:ExpNumVyPsiDistribution}A shows the rheotactic velocity distributions $P(v_y)$ at a given height $z=H/4$ for different local shear rates corresponding to the mean values shown on Fig.~\ref{Fig2:MeanVxVy}C. As expected, without flow and for small shear rates the velocity distribution is very broad and centered around  zero. For increasing shear rates the distributions are shifted more and more towards positive $v_y$ values, together with a small decrease of the width of the distribution.
The experimental and numerical distributions are in excellent agreement and reinforce the validity of our model. Note that using a swimming velocity distribution matching the experimental results is crucial to obtain such an agreement, whereas it is sufficient to work with average values for $\alpha$ and $\nu$.

The rheotactic velocity of each bacterium is a direct consequence of its 3D orientation and its intrinsic swimming velocity $v$ setting the instantaneous velocity $v_y(t)=v e_y=v \cos\Theta(t)\sin\Psi(t)$. Here the angle $\Theta$ is linked to the orientation of the bacterium in the $z$-direction, $e_z = \sin\Theta$, and $e_x=-\cos\Theta\cos\Psi$ and $e_y=\cos\Theta\sin\Psi$ (see Fig.~\ref{SetupTopBottomTracks}B). While bacteria orientations can directly be determined from the numerical simulations, only projections of bacteria trajectories into the $x$-$y$ plane are accessible from the experiments and the in-plane orientations are obtained from the orientation of the velocity vector (see Methods), defined by the angle $\Psi$. 
We thus show the  corresponding orientation distributions $P(\Psi)$ for different local shear rates are shown in  Fig.~\ref{Fig3:ExpNumVyPsiDistribution}B for the layer $z=H/4$. At zero shear rate the orientation distribution is flat and remains so for small shear rates, indicating no preferred bacteria alignment under weak flow.  With increasing shear rate bacteria orient, on average, more and more towards positive values of $\Psi$ and thus towards the right with respect to the negative flow direction. In addition a double peak is emerging, which is found to be mostly symmetric around $\Psi=\pi/2$ corresponding to a bacteria orientation perpendicular to the flow direction.
Bacteria are thus oriented up and downstream around the perpendicular orientation. These peaks become more pronounced and also move closer together with increasing shear rate.
They correspond to a single peak in the velocity distributions as seen on Fig.~\ref{Fig3:ExpNumVyPsiDistribution}A.

To represent a more complete picture of the orientation distributions under flow, Fig.~\ref{Fig4:distributionschannelheight} shows the distributions layer by layer along the channel height.  $z$ scans have been performed at different imposed flow rates $Q$ and are represented as a function of the wall shear rate $\dot{\gamma}_W=4 V_{max}/H$. Higher flow rates are displayed only for the simulation results, as at those large flow velocities the time resolution of the experimental image capture is not sufficient to resolve the bacteria trajectories. The probabilities of the rheotactic velocity and orientation distributions are represented with a color code where yellow corresponds to a high probability and blue to low probability. Peaks in the distributions are thus easily identified as regions of yellow color. For low flow rates, the continuous shift of the peak velocity from zero towards higher velocities when moving away from the middle of the channel towards the channel walls (and thus with increasing local shear rate) is clearly visible, in agreement with Fig.~\ref{Fig2:MeanVxVy}C. For larger flow rates, in layers closer to the channel walls and thus for higher local shear rates only a very weak increase or even a saturation of this peak value can be identified, in particular from the numerical results. 
From the orientation distributions the existence of a double peak is clearly visible for moderate flow rates and a decrease of the distance between the two peaks is observed when getting closer to the channel walls. At high flow rates the simulation results show that the double peak merges into a single, but wide peak.

These figures also show the up- and down-stream orientations induced by the swimming dynamics of the bacteria in the Poiseuille flow, where bacteria cross multiple layers in the $z$-direction during a trajectory \cite{Junot2019, ZottlPRL2012,Ezhilan2015, RusconiNaturePhy2014}. This is in particular visible at small wall shear rates, where in the center of the channel, most bacteria are oriented downstream, as can be seen by the yellow peak close to $\pi$. Closer to the channel walls this peak is found around $0$, corresponding to an upstream orientation, in agreement with the relative bacteria velocities shown in Fig.~\ref{Fig2:MeanVxVy}B.

\subsection{Theoretical understanding of the observed orientation distributions}
To explain the orientation distributions $P(\Psi,\Theta)$ at the origin of the rheotactic behavior we start by  discussing a simple model system for bulk rheotaxis, namely non-tumbling elongated bacteria in linear shear flow with constant shear rate $\dot{\gamma}$.

In the absence of noise the equations for the orientations of chiral microswimmers (Eq.~(\ref{Eq:EOM11})) can be re-written in terms of the angles $\Psi$ and $\Theta$ and read (see also Ref.~\cite{Mathijssen2019}),
\begin{equation}
  \begin{split}
    \dot{\Psi} &=   \frac{\dot{\gamma}}{2} (1+G) \sin\Psi\tan\Theta + \dot{\gamma} \nu \cos\Psi \frac{\cos 2 \Theta }{\cos\Theta}, \\
    \dot{\Theta} &=  \frac{\dot{\gamma}}{2} (1 - G \cos 2 \Theta) \cos\Psi +\dot{\gamma} \nu \sin\Psi\sin \Theta.
    \label{Eq:DETM}
  \end{split}
\end{equation}
Interestingly, both the Jeffery (1st terms) and the rheotacic contributions (2nd terms)  are linear in the shear rate $\dot{\gamma}$ which as a consequence only determines the time scale of the dynamics.

\begin{figure}
  \includegraphics[width=\columnwidth]{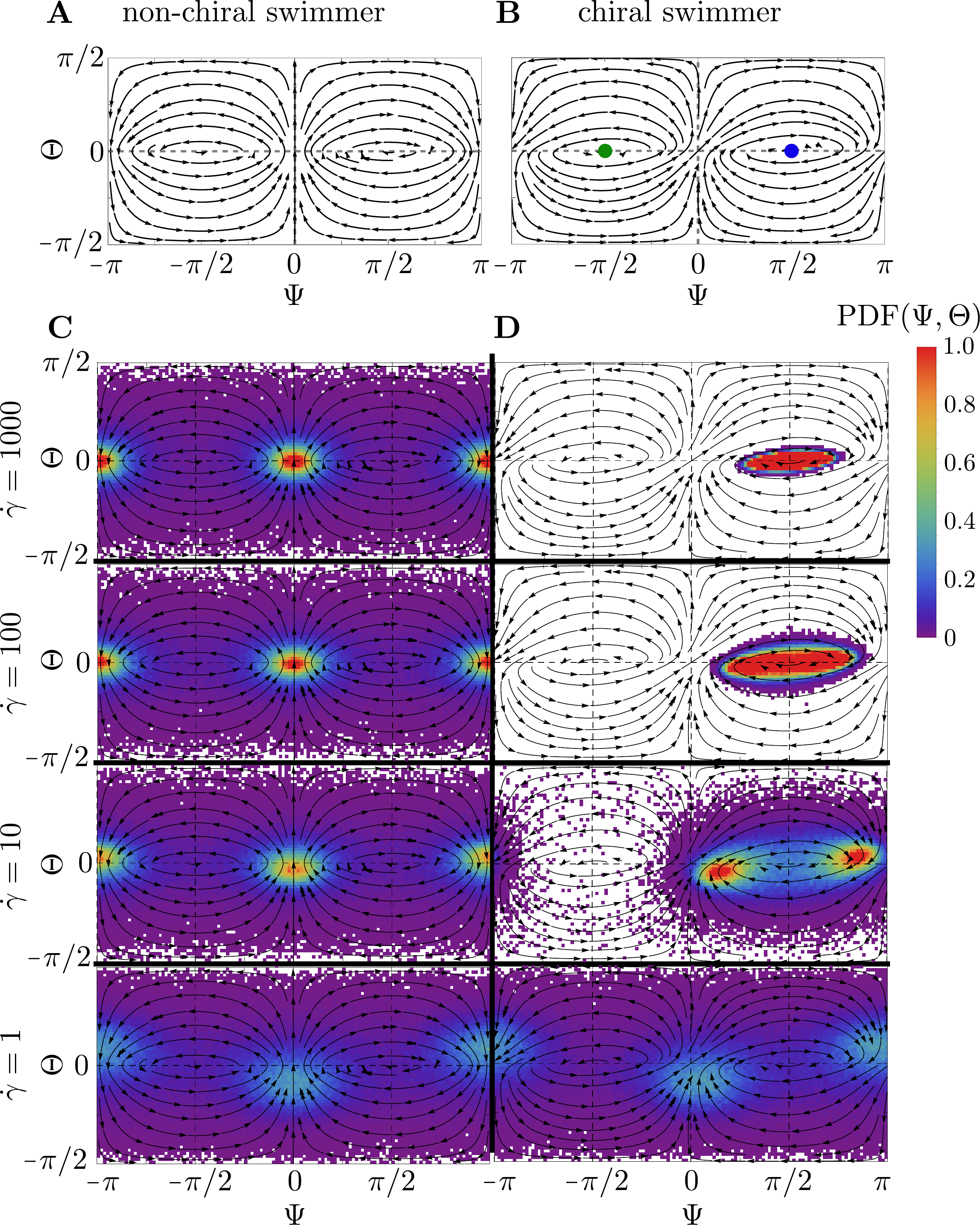}
  \caption{ 
    \textbf{Orientation phase space and simulated probability distributions for non-tumbling bacteria in simple shear flow.}
    \textbf{(\textit{A})} The streamlines of a non-chiral swimmer simply follow Jeffery's periodic solutions for passive ellipsoids.
    \textbf{(\textit{B})} Chirality breaks the left-right symmetry which is the main reason for bacterial bulk rheotaxis. The green and blue dots correspond to marginally stable fixed points in the linear regime pointing to the left ($\Psi=-\pi/2$,$\Theta=0$) and to the right ($\Psi=\pi/2$,$\Theta=0$) side of the channel.
    \textbf{(\textit{C,D})} Orientation distributions for non-chiral \textbf{(\textit{C})} and chiral \textbf{(\textit{D})} swimmers at different shear rates using our standard parameters ($\nu=0.06$, $D_r=0.067s^{-1}$,$\alpha=5$).}
  \label{Fig5:OrientationDist}
\end{figure}

The orientation $(\Psi(t),\Theta(t))$ of non-chiral swimmers ($\nu=0$) simply
follow Jeffery orbits, which are solutions of Jeffery's equations of motion (see also Eq.~(\ref{Eq:OmegaJ})). 
The Jeffery orientation phase space  is shown in  Fig.~\ref{Fig5:OrientationDist}A, and closed streamlines indicate the well-known periodic Jeffery solutions which depend on the initial orientation of the swimmer and correspond to a constant of motion, the so-called Jeffery constant \cite{jeffery1922motion}.
The arrows indicate the direction of motion on the unit sphere. Again, the shear rate $\dot{\gamma}$ does not alter the Jeffery orbits, but only sets the time scale of how fast an elongated particle or bacterium rotates. Note, when a particle starts  oriented to the left/right ($\Psi<0$~/~$\Psi>0$), it never switches to the other side.

Adding the rheotactic term now breaks the left-right symmetry and,
as can be seen in Fig.~\ref{Fig5:OrientationDist}B, all orientations eventually end up on the right ($\Psi>0$) leading to the mean rheotactic velocity $\bar{v}_y > 0$. 
In order to understand the orientational behavior in more detail, we see 
 that  ($\Psi^\ast=\pm \pi/2,\Theta^\ast=0$) are two fixed points of the system [Eqs.~(\ref{Eq:DETM})], defined where
$(\dot{\Psi}^\ast,\dot{\Theta}^\ast)=(0,0)$. 
The eigenvalues of the stability matrix for the linearized system around the fixed points can be evaluated to

\begin{equation}
  \lambda_{1,2} = \pm \frac i 2 \dot{\gamma}  \sqrt{1-G^2-4\nu^2} = \pm i \omega.
  \label{Eq:Theory2}
\end{equation}
Interestingly, these Eigenvalues have a vanishing real part, which means that the fixed points are actually not stable/unstable, but  marginally stable.
In other words, the typical time scale at which a bacterium orients towards the fixed point on the right diverges and it will therefore never be reached.
Indeed we observe that simulated trajectories without noise do not reach the fixed point but get trapped in periodic orbits around the marginally stable fixed point.

We now turn to the question how rotational noise, always present in experiments, influences the orientational distribution of bacteria in flow.  The rotational diffusion constant $D_r$ of the bacterium compared to the strength of the shear rate $\dot{\gamma}$ now plays an important role. For non-chiral swimmers ($\nu=0$) noise effects the dynamics similar as for passive elongated particles
\cite{Saintillan2009}.
Due to noise, trajectories  now erratically move around between different Jeffery orbits \cite{Zoettl2019}  which strongly influences
the  orientation distribution function $P(\Psi,\Theta)$,
as shown in Fig.~\ref{Fig5:OrientationDist}C. Here we show results of 1000 averaged steady-state trajectories started uniformly distributed on the unit sphere for a fixed rotational diffusion $D_r=0.067s^{-1}$ but for a broad range of different shear rates. 
Purple and blue correspond to low probability while yellow and red to high probability (see scale bar). 
The first row  in Fig.~\ref{Fig5:OrientationDist}C shows the case where noise is not yet important, due to the high shear rate $\dot{\gamma}=1000$, 
and $P(\Psi,\Theta)$
almost follows the deterministic solution obtained from Eqs.~(\ref{Eq:DETM}).
The peaks at $(\Psi=0,\Theta=0)$ and $(\Psi=\pi,\Theta=0)$  correspond to particles aligned with the flow, upstream or downstream, respectively.
These peaks are a consequence of the fact that  elongated particles moving on a (typically kayaking) Jeffery orbit have a high probability to come close to these positions, which can be seen by the increased density of streamlines in Fig.~\ref{Fig5:OrientationDist}A.
Together with a decreased rotation rate near these points this leads to the large probability observed. Note that the slowest rotation is near the log-rolling states ($\Psi= \pm \pi/2$) but hardly any initial condition leads to orbits which come close to these states and precludes high probabilities around them.

As can be seen in the second row in  Fig.~\ref{Fig5:OrientationDist}C, reducing the shear rate by a factor 10 does not change
the distributions significantly, meaning that noise is still not important for non-chiral swimmers.
This changes for smaller shear rates,  $\dot{\gamma}=10$, where the peaks of the distributions become less pronounced.
At even smaller flow rates, $\dot{\gamma}=1$ (last row), noise randomizes the orientations almost completely and only very weak peaks remain.
Note that both upstream and downstream peaks are not symmetrically distributed around $\Theta=0$ for small $\dot{\gamma}$ 
due to a subtle interplay between noise and flow 
which leads to nonuniform microswimmer profiles in the $z$-direction \cite{RusconiNaturePhy2014,Ezhilan2015,Saintillan2009}.
In general, the important physical quantity which determines the nonchiral orientation distribution is the rotational P\'eclet number $\text{Pe}_r=\dot{\gamma}/D_r$. As we will see in the following, the situation is less simple when particles are chiral.

For chiral microswimmers ($\nu>0$) the signatures of the marginally stable fixed point can be seen at high shear rates.
As shown in the first row of Fig.~\ref{Fig5:OrientationDist}D, even for very weak noise or high shear rates ($\dot{\gamma}=1000$)
the peak of the distribution is not a simple $\delta$-peak but smeared out around the marginally stable fixed point at $\Psi=\pi/2$.
For  smaller  shear rate  $\dot{\gamma}$, when noise becomes more dominant, the probability distribution becomes even broader, and develops bimodal peaks, which shift more and more towards orientations aligned
(or antialigned) with the flow when lowering $\dot{\gamma}$ further (2nd and 3rd row in Fig.~\ref{Fig5:OrientationDist}D).
The positions of these peaks are triggered by a competition between the chirality-induced attraction to the right and the Jeffery-peaks which become more and more dominant
for smaller shear rates since noise then helps to move around in the phase space more easily.
For low $\dot{\gamma}$ (last row) chiral migration to the right is very inefficient compared to noise, and the probability distributions look very similar to the non-chiral system.

Finally we note that the frequency of oscillation $\omega$ [see Eq.~(\ref{Eq:Theory2})]
is comparable to the  pure Jeffery frequency $\omega_J=\frac{\dot{\gamma}}{2}  \sqrt{1-G^2}$
but is reduced by approximately 5\% for our standard parameters         $\nu=0.06$ and $\alpha=5$.

\begin{figure}[tb]
  \centering
  \includegraphics[width=\columnwidth]{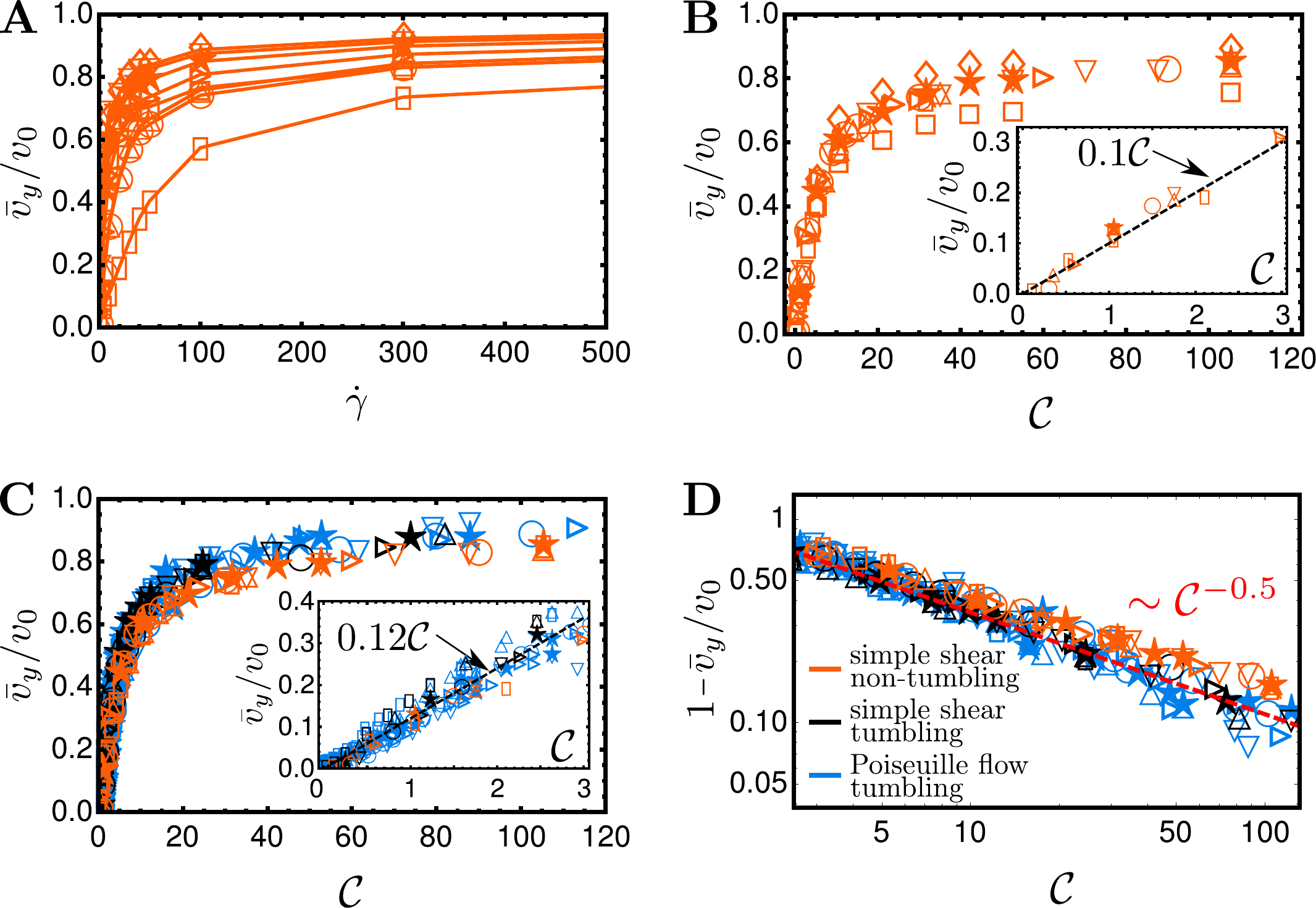}\\
  \caption[test]{ \textbf{Universal scaling of the rheotactic velocity.}
    \textbf{(\textit{A})} Dependence of the scaled mean rheotactic velocity $v_y/v_0$ on the shear rate $\dot{\gamma}$ for non-tumbling bacteria in simple shear (simulations)
    for different parameter sets (rotational diffusion $D_r$, bacterium aspect ratio $\alpha$, chiral strength $\nu$).
    \textbf{(\textit{B})} Results as shown in  \textbf{(\textit{A})} but plotted against the chirality number $\mathcal{C}$. \textbf{(\textit{C})} Data in \textbf{(\textit{B})} for $\alpha=5$ compared to tumbling bacteria in simple shear flow and Poiseuille flow. \textbf{(\textit{D})} Slow algebraic saturation at high shear rates. Color code indicated in \textbf{(\textit{D})}. 
    Symbol code used in all subfigures: \newline
      \includegraphics[height=7pt]{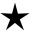} $D_r=0.057$ $\alpha=5$ $\nu=0.06$;
      \includegraphics[height=7pt]{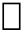} $D_r=0.057$ $\alpha=5$ $\nu=0.006$;
      \includegraphics[height=7pt]{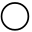} $D_r=0.2$ $\alpha=5$ $\nu=0.06$;
      \includegraphics[height=7pt]{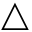} $D_r=0.057$ $\alpha=5$ $\nu=0.02$;
      \includegraphics[height=7pt]{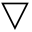} $D_r=0.057$ $\alpha=5$ $\nu=0.1$;
      \includegraphics[height=7pt]{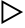} $D_r=0.1$ $\alpha=5$ $\nu=0.06$;
      \includegraphics[height=7pt]{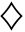} $D_r=0.057$ $\alpha=3$ $\nu=0.06$;
      \includegraphics[height=7pt]{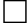} $D_r=0.057$ $\alpha=10$ $\nu=0.06$.
  }
  \label{Fig6:MasterCurve}
\end{figure}

\subsection{Universal scaling and master curve}

We  now analyze the dependence of the rheotactic drift on the system parameters.
Therefore we simulate the dynamics of different rotational diffusion $D_r$, bacterium aspect ratio $\alpha$,
and chiral strength $\nu$ for varying shear rates $\dot{\gamma}$.
Fig.~\ref{Fig6:MasterCurve}A shows the scaled mean rheotactic velocities $\bar{v}_y/v_0$ as a function of the shear rate for different non-tumbling bacteria in simple shear flow.
As expected, none of the curves approach $\bar{v}_y/v_0 \rightarrow 1$ even for very large shear rates because of the aforementioned trapping and oscillations of the swimmer orientations
in the vicinity of the marginally stable fixed point ($\Psi^\ast=\pi/2,\Theta^\ast=0$).

As seen before, for small shear rates all curves increase linearly with the shear rate (Fig.~\ref{Fig2:MeanVxVy}D).
The reason is that for small shear rates
the peaks of the distributions are close to $\Psi^\pm=\{0,\pi\}$ and $\Theta^\pm=0$ (see also Fig.~\ref{Fig3:ExpNumVyPsiDistribution}B), and the peaks linearly shift away from them with increasing shear rate.
The slope is stronger for larger chirality $\nu$ or weaker rotational Brownian noise $D_r$.

In order to understand the behavior at small shear rates
we  linearize Eqs.~(\ref{Eq:DETM}) around  ($\Psi^{\pm},\Theta^\pm$), \textit{i.e.}\ evaluate it at angles $ \Psi=\Psi^{\pm}+\epsilon_\Psi$ and
$ \Theta=\Theta^{\pm}+\epsilon_\Theta$ with small $\epsilon_\Psi$ and $\epsilon_\Theta$.
To lowest order in  $\epsilon_\Psi$ and $\epsilon_\Theta$ we then obtain $\dot{\Psi}(\Psi,\Theta) = \pm \dot{\gamma}\nu$+h.o.t.\ and
$\dot{\Theta}(\Psi,\Theta) = \Omega_\Theta^J$+h.o.t.\ where all higher order terms (``h.o.t.'') are
at least quadratic in the angles $\epsilon_\Psi$ and $\epsilon_\Theta$, and $\Omega_\Theta^J$ is the Jeffery contribution to the angular velocity component $\Omega_\Theta$. 
The reorientation rate $\dot{\Psi}$, which is responsible for the drift towards the right, is to the lowest order independent of $G$ and
linear both in $\dot{\gamma}$ and in $\nu$.
This suggests that the shifts of the peaks $\Psi^\ast$ of the probability distributions -- and hence of the mean values $\bar{v}_y$ -- should be a function
of $\dot{\gamma}\nu$, acting against the noise quantified by $D_r$.

Our analysis, and the results shown in Fig.~\ref{Fig6:MasterCurve}A, now suggest to introduce a dimensionless \textit{chirality number} 
\begin{equation}
  \mathcal{C} = \frac{\dot{\gamma}\nu}{D_r^\text{eff}}
  \label{Eq:C}
\end{equation}
as the main relevant physical quantity which regulates rheotaxis.
Here we use $D_r^\text{eff}$  as the effective rotational diffusion constant
determined via the exponential decay of the bacterium orientation autocorrelation function in the absence of flow, $\langle \mathbf{e}(0)\cdot \mathbf{e}(t) \rangle = e^{-2D_r^\text{eff}t}$ \cite{Zottl2016}.
  It is simply the rotational diffusivity $D_r$ for non-tumbling bacteria
but would be larger for tumbling bacteria due to the extra tumbling reorientations.
In Fig.~\ref{Fig6:MasterCurve}B we now show the same data as in Fig.~\ref{Fig6:MasterCurve}A but as a function of  the chirality number $\mathcal{C}$.
As can be seen in the inset of Fig.~\ref{Fig6:MasterCurve}B, for small shear rates all curves fall on top of each other.
At higher shear rates only curves with the same $\alpha$ (and hence $G$) fall on top of each other.

This result is important as it points in this weak shear regime, on the relevant parameters defining the rheoactic drift velocity for a  microswimmer. The linear relation between mean transverse rheotatic velocity $\bar{v}_y$ and shear rate $\dot{\gamma}$, $\bar{v}_y=l_0\dot{\gamma}$ also reveals a length
  $l_0$ quantifying how far a bacterium drifts in the vorticity direction at unit shear rate. For our bacterial strain this length  is $l_0 \approx 0.75\mu m$ (see Fig.~\ref{Fig2:MeanVxVy}D), and we now know from Eq.~(\ref{Eq:C}) and the prefactors shown in the insets of Fig.~\ref{Fig6:MasterCurve}B,C that $l_0$ is related to the bacteria parameters via $l_0\approx 0.1v_0 \nu/D_r^\text{eff}$.

At larger shear rates, the bacteria with lower aspect ratio ($\alpha=3$, diamonds) have a higher, and the swimmers with higher aspect ratio ($\alpha=10$, squares) a lower rheotactic velocity.
This can be understood by
looking at higher orders of the stability around the locally  marginally stable fixed point around $\Psi=\pi/2$.
Indeed, it can be shown that away from the linearly marginable stable fixed point the attraction towards this point is stronger for smaller $\alpha$, and hence the rheotactic velocity larger.

We also test our scaling for tumbling bacteria, and compare the results to non-tumbling bacteria in Fig.~\ref{Fig6:MasterCurve}C.
Indeed we can see that the scaling works well also for tumbling bacteria, where all of the data (black symbols) collapse onto a single curve.
The non-tumbling and tumbling master curve do not exactly fall on top of each other which suggests that the simple scaling with $D_r^\text{eff}$ is not perfect.
Note, as described in the before, we include tumbling not just as defining an enhanced effective rotational diffusion, but by adding random tumbling events on top of the continous rotational diffusion.
While in the absence of flow long-time dynamics may be captured by introducing an effective temperature from steming from an effective diffusion constant \cite{Zottl2016},
this concept fails to be the accurate measure in our situation, where the interplay of flow and tumbling prohibits a simple effective temparature scaling.
Again we obtain a linear regime for small $\dot{\gamma}$ (inset) where the slope is almost the same (but not exactly) as for the non-tumbling bacteria.

Now we turn to the case of tumbling bacteria  in Poiseuille flow by plotting the data extracted from different layers in the channel (and hence local shear rates),
as shown in blue in Fig.~\ref{Fig6:MasterCurve}C.
The fact that this data collapses together with the simple-shear data on a single curve reiterates the fact that, at least for our channel height ($H=100\mu m$),  bacterial rheotaxis is 
determined by the local shear rate.

Finally we show in  Fig.~\ref{Fig6:MasterCurve}D how the curves in Fig.~\ref{Fig6:MasterCurve}C approach towards their asymptoptic values.
As discussed before, bacteria approach the infinite-shear limit very slowly, which can be fitted, as a first approximation, to a power law $\sim\mathcal{C}^{-0.5}$,
as a consequence of the marginally stable fixed point at $\Psi=\pi/2$.

\subsection{Comparison to experimental and  numerical results}

Our theoretical analysis allows us now to understand all the observed aspects of the rheotactic velocities in both experiment and Brownian dynamics simulations.

First, 
the aforementioned result  that the position of the peaks in $P(\Psi)$ for chiral bacteria move closer and closer together with increasing shear rate (Fig.~\ref{Fig5:OrientationDist}D)
is clearly observed in the experiments and simulations in Poiseuille flow using tumbling bacteria, as shown for $P(\Psi)$  in Figs.\ref{Fig3:ExpNumVyPsiDistribution}~B and \ref{Fig4:distributionschannelheight}~B.

Second, we can now understand why the plateau value of the mean rheotactic velocities obtained in experiments and simulations for high shear rates does not approch the mean free swimming speed $v_0$ (Fig.~\ref{Fig2:MeanVxVy}C): The occurrence of a marginally stable fixed point traps bacteria in periodic-like orbits  resulting in broad orientational distributions.

Third, we are able to predict the linear increase of the mean rheotactic velocity for small shear rates as seen experimentally and in simulations in Fig.~\ref{Fig2:MeanVxVy}C, and to determine the dependence of the mean rheotactic velocities on the relevant parameters resulting in a universal scaling law.

Forth, we demonstrate that 
our analysis in simple shear flow qualitatively explains the behavior in Poiseuille flow.
We find that the orientatation probability distributions $P(\Psi,\Theta)$
in Poiseuille flow obtained from layers  with local shear rate $\dot{\gamma}$ follow
the respective distributions in simple shear flow.
Tumbling allows bacteria to exploit non-popular regions in phase space more frequently compared to non-tumbling bacteria, but the qualitative results are comparable to the ones of  non-tumbling bacteria.

\section{Conclusions}
In this paper we investigated bacterial rheotaxis in bulk flows using a combined experimental, numerical and theoretical analysis. Precise microfluidic experiments using \textit{E.~coli} bacteria in channel flows, provide not only average rheotactic velocities in the vorticity direction as a function of local shear rate in the Poiseuille flow, but also accurate velocity and orientation distributions. These results are in perfect agreement with Brownian dynamics simulations and indicate a shift of the peaks of the velocity distributions towards increasing rheotactic velocities with increasing shear rates. While these peaks tend towards a maximum rheotactic velocity equal to the bacteria swimming speed and corresponding to a bacteria orientation perpendicular to the flow direction and thus aligned with the vorticity direction, the velocity distributions remain very large and average rheotactic velocities always remain significantly below the bacteria swimming speed. 

By theoretically analyzing the bacteria orientation dynamics, we elucidate the mechanisms at the origin of the rheotactic behavior and show comprehensively how the interplay between Jeffery orbits, rheotactic torque and noise leads to the observed rheotactic velocities. Chirality and bacteria geometrical features are  encoded into a dimensionless chiral strength which can be used together with a rotational P\'eclet number to rescale the shear rate. At small shear rate the resulting chirality number affects proportionally the mean rheotactic drift velocity.
Such a regime exists only in the presence of a strong stochastic reorientation process limiting the natural tendency for the chirality-induced torque
to reorient
the swimmer in the positive vorticity direction.

When stochasticity is less important \textit{i.e.}\ at higher chirality number or shear rate, the complexity of the dynamical processes will play a central role and limits in  a subtle way the alignment of the microswimmers perpendicular to the flow. Our theoretical analysis based on a full set of kinematic equations provides a quantitative account for this original dynamical behavior and explains the reorientation dynamics leading to a saturation at large shear rates from the marginal stability of a fixed point.

Our analytical model is not specific to \textit{E.~coli} bacteria and can in the future also be used to describe rheotaxis of other flagellated microorganisms or even a rheotactic torque of a different nature as might exist for artificial microswimmers. It might also be extended to more complex flow environments as soils or porous media and could open interesting perspectives for the design of separation or filtration devices.

\section{Materials and Methods}
\setcounter{figure}{0}
\renewcommand{\thefigure}{M\arabic{figure}}

\subsection{Bacterial cultures}
	
We use the wild type bacteria strain RP437 with fluorescently stained cell bodies, emitting green fluorescence light. These bacteria perform run and tumble motion with a typical tumbling frequency of 1 Hz. A rich growth medium (M9G) is used for the bacterial culture and is prepared from 5.64g M9 salt, 2g glucose, 0.5g casamino acid, 50mL CaCl2 (1M), 1mL MgSO4 (1M) in pure water in a total volume of 500mL.	Culturing with antibiotics (Chloramphenicol, $25\mu g/mL$) is performed in the oven at 30$^{\circ}$C at a shaking rate of 200 rpm for about 16 hours until an optical density of $OD\sim0.5$ (measured by Eppendorf D30 at $\lambda=600nm$) is reached. The uniformity of the bacteria shape and bacteria mobility are verified under the microscope and subsequently the bacterial suspension is centrifuged at $5000$rpm for 5 minutes and re-dispersed into a motility buffer (0.1mM EDTA, 1$\mu M$ L-methionine, 10mM sodium Lactate, and potassium phosphate buffer with 0.01$M$ at ph=7) containing also L-Serine at a concentration of 0.04$g/mL$.  Bacteria and motility buffer are density matched by adding Percoll (23\% w/w colloidal silica particles of 15-30 nm diameter in water, coated with polyvinylpyrrolidone) at a 1:1 volume ratio. We work with a very small final density of around 0.15~$\% V/V$ (corresponding to an optical density of $OD=1.8$).

\subsection{Microfluidic devices}
	
Rectangular PDMS channels with the height $H = 100\mu m$, width $W = 600\mu m$ and length of 20$mm$ is fabricated using standard soft lithography techniques. A small PDMS layer is spin-coated onto the bottom glass slide to obtain full PDMS channel walls, and also to avoid bacteria sticking. 	
A high precision syringe pump (Cellix ExiGo, Ireland) is used to introduce the bacterial suspension into the microchannel at well controlled flow rates $Q=0, 1, 2, 4, 8, 15, 20, 30, 40$ and $50nL/s$, corresponding to wall shear rates of $\dot{\gamma}_{wall}=0, 1, 2, 4, 8, 15, 20, 30, 40, 50s^{-1}$.

\subsection{Microscope visualization}
	
The bacterial suspensions are visualized using an inverted microscope (Zeiss-Observer, Z1), with an air objective (63x/0.75 LD Plan) and equipped with a Hamamatsu camera (ORCA-Flash 4.0, C11440) at a frame rate of 200fps at $1024\times 512$ pixels (typical field of view size $200\mu m \times 100\mu m$). Using a high frame rate is important to track bacteria at high flow rates, since bacteria displacements in between two frames need to be small compared to a typical distance between two adjacent bacteria.  Due to the use of an air lens there is a mismatch of refraction index with the solution in the channel and height measurements need to be corrected by a factor 1.3622. Two methods are employed to control the local shear rates in the channel: varying the flow rate $Q$ at a given distance from the bottom wall ($z=0.1H$ and $z=0.2H$), called Q-scan and  gradually increasing the distance from the bottom wall with steps $\delta z=5\times 1.3622\approx 6.8\mu m$ at the given flow rates $Q=5, 10, 20nL/s$, called $z$-scan. For each position at a given flow rate, 2000 frames are taken as one stack video for the following tracking process. The fluorescent intensity of the RP 437 strain is sufficient to allow for a small exposure time of 3$ms$.

\subsection{Tracking and analysis} 

Passive polystyrene beads ($1\mu m$ diameter, emitting red fluorescence light) are mixed with the bacterial suspension at very low concentration (smaller than the bacteria concentration) and are introduced together into the PDMS channel. Bacteria and bead trajectories can now be recorded during the same experiment using either a red or a green filter at a frame rate of 200fps.  Considering the mean speed of bacteria of $\sim 25\mu m/s$, and a tumbling event about every second, the positions detected are used at time steps of $\delta t = 0.1s$ (corresponding to every 20 frames) to determine instantaneous bacteria velocities and orientations. 

The depth of field of the used lens has been checked experimentally to be around 2$\mu m$ and observations thus take place within a fluid layer of this thickness. During $\delta t = 0.1s$  bacteria can typically not displace over distances larger than the layer thickness. Hence, we do not filter specific trajectories that are oriented preferentially parallel to the observation plan, but capture all trajectories independently of their orientation. The measured trajectories then represent projections of 3D trajectories into the plane of observation. 
	
From the original video stack (2000 frames for each), using the TrackMate routine (Marcro in FIJI) the positions of individual bacteria are identified and linked to smooth trajectories. The main parameters used for the tracking routine include Laplacian of Gaussian detector, a blob diameter of 5 pixels, 
LAP tracker (with maximum distance varying as a function of the flow rate). After detecting and linking the spots for every bacteria, all the trajectories are saved for the later extraction of the positions [$x^i(t),y^i(t)$] from the $i^{th}$ trajectory. Similar, for the passive beads the positions [$X^i(t),Y^i(t)$] are measured, and the mean bead velocity in a specific $z-$layer is obtained to $(V_x(z),V_y(z))=\langle (X^i(t+\delta t)-X^i(t),Y^i(t+\delta t)-Y^i(t))/ \delta t   \rangle$ where we average over trajectories and time.

For two adjacent position pairs $[x^i(t),y^i(t)], [x^i(t+\delta t),y^i(t+\delta t)]$
of the $i^{th}$ trajectory, the velocity in the $x$-direction  in a given $z$-layer (composed by the swimming and the flow velocity) at time $t$ as $v_x^i(t) =\frac{ [x^i(t+\delta t)-x^i(t)]}{\delta t}$, and the rheotactic velocity $v_y^i(t) =\frac{ [y^i(t+\delta t)-y^i(t)]}{\delta t}$ is calculated. The unit vector of the bacteria for this $i^{th}$ trajectory at the time $t$ is defined as
$\mathbf{e}_{2D}^i(t)=(\hat{v}_x^i(t),\hat{v}_y^i(t))=((v_x^i(t)-V_x(z))/v_{2D}^i(t),v_y^i(t)/v_{2D}^i(t))$ with $v_{2D}^i(t)=\sqrt{(v_x^i-V_x(z))^2(t)+(v_y^i)^2(t)}$
  where the background flow has been subtracted. The instantaneous orientation angle is then defined as $\Psi^i(t)=\arctan(\hat{v}_y^i(t)/\hat{v}_x^i(t))$.

\subsection{Theoretical model and numerical methods}
We approximate the shape of a swimming \textit{E.~coli} bacterium by an active, ``chiral'' ellipsoid
of 1 $\mu m$ width and $\alpha$ $\mu m$ length, where $\alpha$ is the aspect ratio,
and chiral strength $\nu$.
In addition to rotational diffusion, tumbling bacteria tumble at exponetially distributed tumble times $\sim exp(-t/ \tau)$ with $\tau=1s$.
This is performed by an instantaneous rotation about a random axis around a random angle $\beta$ drawn from a Gaussian distribution  with mean $\beta_0=1.082 rad$ and standard deviation $\delta\beta=0.454rad$, in accordance with experimentally observed tumbling statistics \cite{berg1972chemotaxis}.

The equations of motion of a bacterium in flow is given by Eqs.~(\ref{Eq:EOM11a}) and (\ref{Eq:EOM11}). 
$\mathcal{H}$
is calculated from the translational  diffusion tensor $\textbf{\textsf{D}}(\Psi,\Theta)=\bar{D}{\textbf{\textsf{1}}} + \frac 1 2 \Delta D \textbf{\textsf{M}}(\Psi,\Theta) =\frac 1 2 \mathcal{H} \cdot \mathcal{H}^T$
via Cholesky decomposition
where $\textbf{\textsf{M}}(\Psi,\Theta)$ is a symmetric 3x3 matrix (see \cite{Matsunaga2017}),
and $\bar{D} = (D_1 + D_2)/2$, $\Delta D = D_1 - D_2$ where $D_1=k_BT a^{-1} \eta^{-1} K_1(\alpha)$ and $D_2 = k_BT a^{-1} \eta^{-1} K_2(\alpha)$
are the respective longitudinal and transversal diffusion coefficients of an ellipsoid of aspect ratio $\alpha$ with shape functions $K_1(\alpha) > K_2(\alpha)$ (see \cite{Matsunaga2017, Zoettl2019}),
and with the effective particle (bacterium) radius $a=\sqrt[3]{3V_p/(4\pi)}$ where $V_p$ is the volume of the particle (bacterium).
We use room temperature hence $k_BT=4.14pNnm$ and buffer viscosity $\eta=1.28\times 10^{-3}Pas$.
The random numbers $\xi_i$ and $\xi_i^r$ model Gaussian white noise with zero mean and $\langle \xi_i \xi_j \rangle=\langle \xi_i^r \xi_j^r \rangle=\delta_{ij}$ ($i=x,y,z$).

In order to compare results with the experiments, we determine the instantaneous velocity $\mathbf{v}=v_x\hat{\mathbf{x}} + v_y\hat{\mathbf{y}} +v_z\hat{\mathbf{z}}$ of the swimmer at time $t$ by using
$\mathbf{v}= [\mathbf{r}(t+\Delta t) - \mathbf{r}(t)] / \Delta t$ with $\Delta t = 0.1s$, similar as in the experiments.
      
%\section*{Supplementary Materials}
%\noindent Supplementary Text \\
%Fig.~S1: Velocity distribution in the absence of flow.
	
\section*{acknowledgments}
GJ gratefully acknowledge the support by NSFC (No.~11774287).
AL and AZ acknowledge funding from the ERC Consolidator Grant PaDyFlow (Agreement 682367).
AZ acknowledges funding from the Austrian Science Fund (FWF) through a Lise-Meitner Fellowship (Grant No.~M 2458-N36).
EC and AL acknowledge funding from Agence Nationale de la Recherche ``Bacflow'' Grant ANR-15-CE30-0013.
This work received the support of Institut Pierre-Gilles de Gennes (\'Equipement d'Excellence, ``Investissements d'Avenir", Program ANR-10-EQPX-34). \\

\section*{Author contributions}
GJ, AZ, AL and EC designed the research. GJ performed the experiments and GJ and EC performed data analysis. AZ performed theoretical analysis and numerical modeling. GJ, AL and AZ wrote the manuscript.  All authors contributed to the analysis and interpretation of the results.

%\section*{Competing interests}
%The authors declare that they have no competing interests.

%\section*{Data and materials availability}
%All data needed to evaluate the
%conclusions in the paper are present in the paper and/or the Supplementary Materials.
%Additional data related to this paper may be requested from the authors.

\bibliography{RefRheotaxisRef}

\end{document}